\documentclass[12pt]{spieman}
\usepackage{amsmath,amsfonts,amssymb}
\usepackage{graphicx}
\usepackage{setspace}
\usepackage{tocloft}
\usepackage{upgreek}
\usepackage{multirow}%
\usepackage{caption}

\title{Nanosecond-latency all-optical fiber sensing with in-sensor computing}

\author[1]{Yu Tao}
\author[1,*]{Yangyang Wan}
\author[1]{Ziwen Long}
\author[1]{Wenjia Zhang}
\author[1]{Jiangbing Du}
\author[1]{Zuyuan He}

\affil[1]{State Key Laboratory of Photonics and Communications, Shanghai Jiao Tong University, Shanghai 200240, China.}

\cftpagenumbersoff{figure}
\cftpagenumbersoff{table} 
\begin{document} 
\maketitle

\begin{abstract}
Optical fiber sensing plays a crucial role in modern measurement systems and holds significant promise for a wide range of applications. 
This potential, though, has been fundamentally constrained by the intrinsic latency and power limitations associated with electronic signal processing.
Here, we propose an all-optical fiber sensing architecture with in-sensor computing (AOFS-IC) that achieves fully optical-domain sensing signal demodulation at the speed of light. 
By integrating a scattering medium with an optimized diffractive optical network, AOFS-IC enables linear mapping of physical perturbations to detected intensity, and sensing results can be directly read out without electronic processing.
The proposed system maintains high accuracy across various sensing tasks, providing sub-nano strain resolution and 100\% torsional angle classification accuracy, as well as multiplexed sensing of multiple physical quantities, and performing multi-degree-of-freedom robot arm monitoring.
AOFS-IC eliminates computing hardware requirements while providing \textless 3 ns demodulation delay, which is more than 2 orders of magnitude faster than conventional fiber optic sensing systems.
This work demonstrates the potential of next-generation optical sensing systems empowered by all-optical computing, and paves the way for expanded applications of fiber sensing through the integration of fully optical components, ultrafast measurement speed, and low power consumption.

\end{abstract}


{\noindent \footnotesize\textbf{*}Yangyang Wan,  \linkable{YangyangWan@sjtu.edu.cn} }

\begin{spacing}{1}   

\section{Introduction}
Optical fiber sensors have provided valuable assistance in sensing and measurement, with distinct advantages over traditional sensors in terms of sensitivity, spatial resolution, compactness, and immunity to electromagnetic interference\cite{elsherif2022optical, pendao2022optical}. 
The widespread deployment of optical fiber sensors in applications such as seismic monitoring\cite{lindsey2021fiber}, transportation infrastructure\cite{du2020review, wu2020recent}, energy safety\cite{ashry2022review, yi2023sensing}, and precision manufacturing\cite{shimizu2021insight, zhou2023fiber} has led to a growing demand for large-scale and multidimensional optical sensing systems with low latency and high energy efficiency.
Traditional optical fiber sensing (OFS) architecture typically separate the sensing unit from the computing unit, transmitting massive amounts of raw data to centralized processors after optoelectronic conversion, where it undergoes demodulation and analysis. 
While this approach ensures general applicability, it burdens computational resources, increases power consumption, and introduces significant latency, particularly in multiplexed sensor arrays\cite{venketeswaran2022recent}. 
There is a need to develop novel low-power and low-latency fiber-optic sensing architectures to address the challenges posed by future large-scale, high-density deployments of optical fiber sensors\cite{gherlone2018shape, zhu2022structural}.

Many efforts try to achieve low delay or low power consumption by employing advanced algorithms and hardware optimizations.
Although neural network algorithms have been introduced to improve data processing performance\cite{wang2019deep, li2021dilated, wang2025high}, their implementation relies on electrical computing resources.
Field programmable gate arrays (FPGAs), graphics processing units (GPUs), and micro control units (MCUs) are also used to accelerate the demodulation process of optical fiber sensors\cite{hu2018high, cheung2023parallel, wang2024real}.
These methods remain constrained by conventional optical fiber sensor architecture, fundamentally reliant on electrical computing hardware that introduces inherent computational bottlenecks, resulting in persistent difficulties in achieving sub-microsecond latency thresholds for OFS signal processing\cite{venketeswaran2022recent}.

Optical computing has emerged as a promising alternative to conventional electronics, offering intrinsic advantages such as ultra-fast parallel processing, sub-nanosecond latency, and extremely low energy consumption\cite{mcmahon2023physics}.
For example, the designed diffraction surfaces can be regarded as optical diffraction network, which enables feature extraction\cite{zhang2024memory, xia2024nonlinear}, classification\cite{oguz2024programming, xue2024fully, wang2024opto, guan2025photon}, and reconstruction of images\cite{luo2022computational, mengu2022all, wang2023image, yan2024nanowatt, yu2025all} without electrical computing resources.
Diffraction optical neural networks can process data at the speed of light and provide unprecedented parallel processing capabilities to manipulate synthesized complex optical fields\cite{hu2024diffractive}.
These advances indicate that optical computing may bring transformative benefits of high-speed and low power consumption to OFS.
OFS systems usually convert the measured physical quantities into measurements of certain optical field dimensional parameters, such as wavelength, polarization, and intensity. 
Taking a typical commercial fiber Bragg grating (FBG) as an example, its wavelength measurement accuracy should be at least 10 pm\cite{li2021fbg}, which is difficult to achieve for existing optical nerual network (ONN) implementations.
To date, an all-optical fiber sensing system with competitive sensing performance and no electrical processing resources remains largely unexplored.

Here, we propose an all-optical fiber sensing architecture with in-sensor computing (AOFS-IC) that achieves light speed signal processing without electronic processors.
The proposed architecture realizes signal demodulation of OFS in the optical domain, establishing output signal intensity to measured physical quantity mapping via an optical computing system.
The basic principle relies on a scattering medium sensitive to optical frequency, polarization, and spatial modes to encode sensing information into speckle patterns\cite{kohlgraf2010transmission}, with subsequent optimized diffractive optical computing module decoding these patterns into output light intensity.
Instead of  demodulation or reconstruction algorithms, the sensing information can be directly readout from photodetector-measured light intensity, enabling low-power consumption with ultra-low latency.
By using a multimode fiber (MMF) as the scattering medium in AOFS-IC, the accurate measurement of strain in the FBG based OFS is achieved.
Besides, AOFS-IC has also been verified in OFS systems using single-mode fiber (SMF) or MMF as the sensor, and the quantitative measurement of torsion angle, stretch or vibration has been realized with a delay of \textless 3 ns.
By tailoring scattering medium and diffractive optical computing module, AOFS-IC can be flexibly designed to directly read out the physical quantity of interest in specific sensing tasks with tunable accuracy and dynamic range.
Moreover, we demonstrate that AOFS-IC supports multiplexing of multiple fiber-optic sensors while enabling simultaneous direct sensing of diverse physical quantities.

\section{Results}

\subsection{Principle of AOFS-IC}
\begin{figure}[htbp]
	\centering
	\includegraphics[width=1\linewidth]{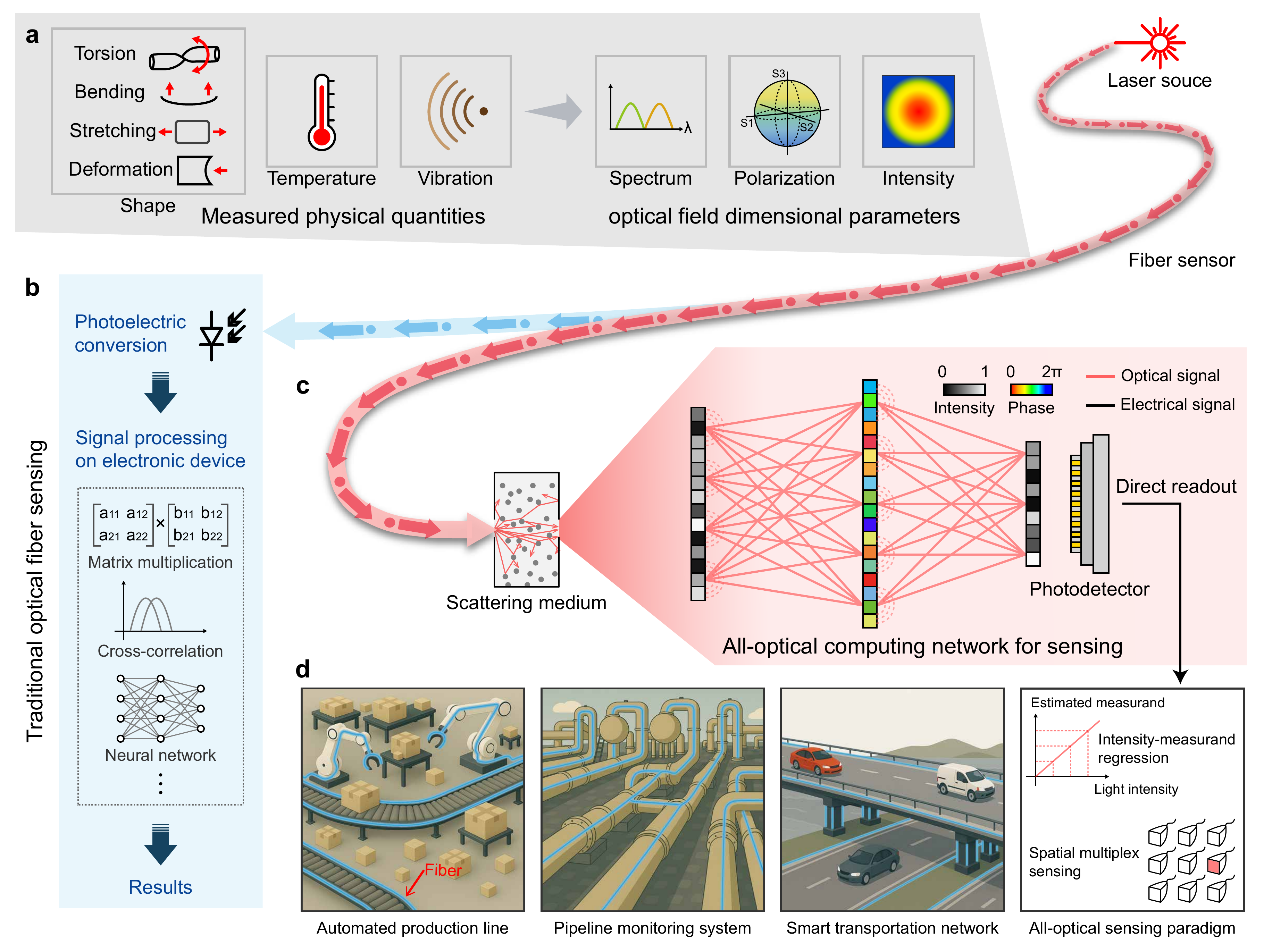}
	\caption{
		\textbf{All-optical fiber sensing architecture with in-sensor computing (AOFS-IC).} \textbf{(a)} optical fiber sensor detects physical quantities including bending, stretching, deformation, vibration, and temperature, with all variations ultimately manifesting through changes in optical field dimensional parameters such as spectrum, polarization state, and intensity distribution. \textbf{(b)} Conventional optical fiber sensing architecture, where the optical signals detected by the fiber are converted into the electrical signals and processed by time-consuming digital algorithms to extract the target measurand on electronic computing devices. \textbf{(c)} The all-optical computing module of AOFS-IC performs optical-domain demodulation of sensing signals, enabling direct quantification of measured quantity through detected intensity from photodetector. The module comprises a scattering medium coupled with an optical diffractive network. \textbf{(d)} AOFS-IC has the advantages of low hardware resource requirements and low latency, and can support sensor multiplexing through spatial multiplexing, which makes it a promising solution for applications in automated production lines, pipeline monitoring systems, and intelligent transportation networks.
		} 
	\label{Arch}
\end{figure}

OFS detects external physical quantities by monitoring changes in optical field parameters, such as spectrum, polarization and intensity, during lightwave propagation through the fiber sensor, as shown in Fig.~\ref{Arch}(a).
This enables precise measurements of temperature, pressure, strain, vibrations, and other physical quantities.
In the traditional OFS architecture, digital signal processing following optoelectronic conversion is typically required to extract physical quantities encoded in the optical field from received signals, introducing inherent latency and electronic computational demands on electronic resources, as shown in Fig.~\ref{Arch}(b).

Fig.~\ref{Arch}(c) demonstrates the proposed AOFS-IC, where optical signal carrying sensing information emitted from the fiber sensor is directly passed through an all-optical computing module for extraction of measured physical quantities without digital signal processing.
The all-optical computing module consists of a scattering medium and an optical diffraction network (ODN).
Since the scattering medium is sensitive to the change of optical field dimension parameters, the subtle optical field changes can be amplified to significant speckle intensity variations through high-dimensional spatial projection\cite{yuan2023geometric}.
The speckle pattern after the scattering medium exhibits dynamic evolution correlated with variations in the physical quantity\cite{redding2014high}.
This evolution is typically quantified by the cross-correlation between speckle patterns, with larger variations in the physical quantity leading to faster decorrelation (see Supplementary Note~3 and Fig.~4).
Although the speckle cannot directly represent the sensing information, it can be decoded and demodulated at the speed of light within an ODN.
Through end-to-end training and optimization, a designed ODN can establish a linear regression relationship between an physical quantity of interest and the output light intensity at the designated region.
Since the demodulation of sensing signal is completed in the optical domain, the physical quantity can be estimated directly from the light intensity received by the photodetector (PD), achieving ultrahigh-speed sensing capability.
For densely multiplexed optical fiber sensors, AOFS-IC engineers the ODN to spatially focus individual sensing signals onto segregated positions, coupled with a detector array for parallel acquisition.
AOFS-IC shows potential for large-scale monitoring applications such as mechanical deformation monitoring, structural health monitoring and pipeline route monitoring, enabled by its low-latency operation, reduced electric computational demands, and multiplexing support, as shown in Fig.~\ref{Arch}(d).

In contrast to conventional diffractive optical computing approaches, our system employs a scattering medium to enhance the sensitivity of ODN to minute optical field variations, so as to meet the accuracy requirements for OFS applications.
In the implementation of AOFS-IC, MMF is selected as the scattering medium. 
The reason is that MMF can generate speckle patterns sensitive to changes of optical dimensional parameters through multimode interference among numerous transmission modes, while maintaining low insertion loss via direct coupling with sensing fiber and achieving compact size through coiling\cite{redding2014high}.
The optical computing network utilizes the programmable spatial light modulator (SLM) due to its flexible and adjustable characteristics for demonstration, which can alternatively be replaced by several passive phase plates (see Supplementary Note~8) to further improve energy efficiency in practical applications.

\subsection{Linear regression between measurand and output intensity}
We first validate AOFS-IC using a classical FBG sensor, which detects physical quantities through spectral shifts.
Standard FBG applied strain induces a linear shift of the Bragg wavelength with the strain sensitivity of 1.2 pm/$\upmu\upvarepsilon$.
In conventional architectures, wavelength shift measurements in such spectrally responsive optical fiber sensors typically rely on optical spectrum analyzers (OSAs) or interrogator-based techniques\cite{li2021fbg}.
Commercial spectrometers typically achieve resolutions no better than 0.01 nm, while interrogation techniques such as Pound-Drever-Hall (PDH) locking involve complex optical configurations and demodulation processes, leading to inherent electronic processing delays and substantial hardware power consumption in conventional approaches.

\begin{figure}[htbp]
	\centering
	\includegraphics[width=1\linewidth]{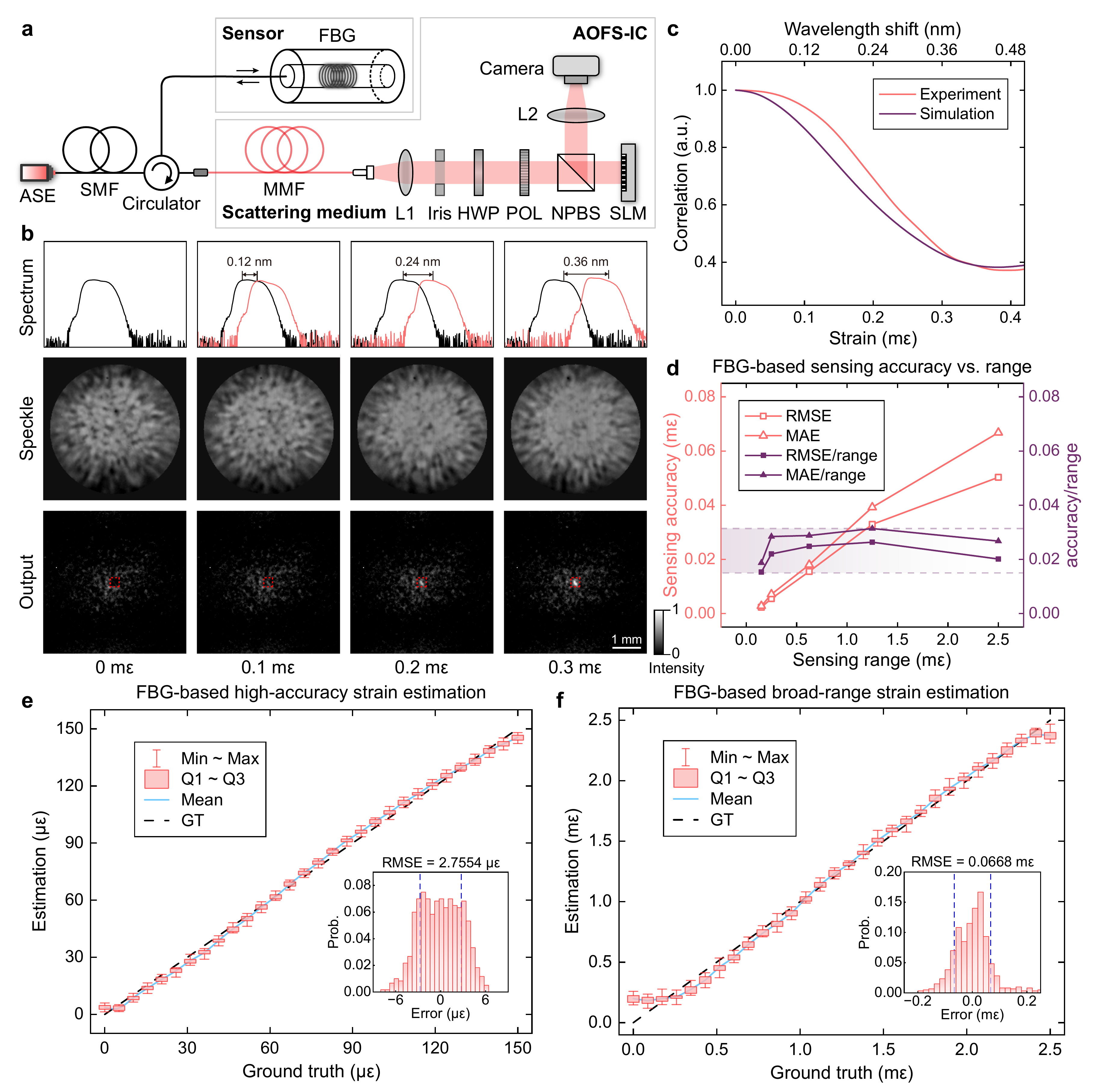}
	\caption{
		\textbf{All-optical estimation of strain over varying dynamic ranges.} \textbf{(a)} Experimental setup of all-optical sensing system based on FBG. The FBG detects variations in strain as the sensing element. \textbf{(b)} Reflection spectrum of FBG under different strain conditions, corresponding speckle patterns after transmission through the MMF, and the resulting light intensity outputs from the optical computing module. \textbf{(c)} Correlation between speckle patterns as a function of applied strain or wavelength shift. Both experimental and simulated results exhibit a gradual decorrelation trend with increasing strain. \textbf{(d)} Trade-off between sensing accuracy and sensing range, where larger sensing ranges correspond to reduced accuracy. RMSE and mean absolute error (MAE) are used to quantify sensing resolution. Normalized accuracy metrics (RMSE/range and MAE/range) are used to evaluate the relative error levels under different perceptual ranges. \textbf{(e--f)} Strain estimation performance with high accuracy (e) and broad measurement range (f), evaluated through measurement of 30 continuous strain states to reflect overall performance. For each strain state, 20 repeated measurements are recorded and statistically analyzed, including the minimum and maximum (Min--Max), first and third quartiles (Q1--Q3), mean, and ground truth (GT). The subplots show histograms of estimation errors, yielding an RMSE of 0.0688 m$\upvarepsilon$ over a 2.5 m$\upvarepsilon$ range, and 2.7554 $\upmu\upvarepsilon$ over a 150 $\upmu\upvarepsilon$ range, respectively. ASE, amplified spontaneous emission; L1/L2, lenses; HWP, half-wave plate; POL, linear polarizer; NPBS, non-polarizing beam splitter.
	} 
	\label{FBG}
\end{figure}

In AOFS-IC, the optical signal reflected by the FBG sensor directly enters the optical computing module through a circulator as shown in Fig.~\ref{FBG}(a).
MMF is used as scattering medium and connected with SMF in circulator by offset fusion splicing to excite a sufficient number of transmission modes.
The reflection spectrum of a FBG shifts in response to variations in applied strain, and the measurement results by an OSA are shown in the top row of Fig.~\ref{FBG}(b).
Optical fields with different spectral information produce uniquely distinguishable speckle patterns after propagating through the MMF, and the speckles measured by a camera are shown in the second row of Fig.~\ref{FBG}(b).
As the applied strain varies, the resulting speckle patterns decorrelate in a deterministic manner as shown in Fig.~\ref{FBG}(c), with the degree of decorrelation proportional to the magnitude of wavelength shift (as well as the strain). 
This primarily arises from wavelength-dependent optical path differences in the MMF, leading to distinct speckle interference patterns. 
To establish linear regression between the estimated strain and the output optical intensity, we train an ODN to perform all-optical transformation of the speckle patterns. 
The training process of ODN employs a designed genetic algorithm\cite{michalewicz1996genetic} integrated with a limited set of strain states (see Supplementary Note~6).
The trained ODN comprises a phase modulation layer that maps directly the high-dimensional speckle field to the output intensity of a designated region, where the total intensity maintains a linear relationship with the applied strain.
As illustrated in Fig.~\ref{FBG}(b), the designated region exhibits progressive enhancement in signal intensity with increasing applied strain from 0 m$\upvarepsilon$ to 0.3 m$\upvarepsilon$.
Due to the relatively broad reflection spectrum of the FBG ($\sim$0.3 nm), spectral overlap may occur between different central wavelengths, resulting in blurred speckle patterns at the MMF output. This speckle ambiguity adversely affects the final sensing resolution of the system. Employing FBGs with narrower bandwidths can mitigate this performance degradation. Despite the possible spectral overlap limit of 0.3 nm, our optical computing system also accurately identifies the wavelength shifts as small as 3.3 pm through the subtle speckle pattern evolution.

As shown in Fig.~\ref{FBG}(e), the applied strain values demonstrate a linear relationship with the intensity of PD, i.e., the estimated strain. 
During the training of the optical computing module, only a minimal number of strain states (e.g., four) are employed.
Notably, despite the optical computing module never encountering the vast majority of strain states during training, it can still map them linearly to optical intensity. 
This capability arises from the inherent memory effect\cite{li2021memory} in the scattering medium, where the speckle pattern gradually decorrelates as a certain parameter changes, as shown in Fig.~\ref{FBG}(c). 
This progressive decorrelation process enables the ODN to approximate linear mapping of speckle variations using only a limited number of states (see Supplementary Note~2).
Under 20 repetitions of the experimental measurements, the strain resolution of the system is statistically defined as the root mean square error (RMSE) between estimated results and real strain. The resolution of the system is 2.7754 $\upmu\upvarepsilon$ (1.85\% of the strain range) over a strain range of 150 $\upmu\upvarepsilon$ over multiple trials.
AOFS-IC enables flexible measurement reconfiguration by simply adjusting the ODN to meet different application requirements.
Fig.~\ref{FBG}(f) demonstrates the enhanced large-strain measurement results.
Even for dynamic ranges as large as 2.5 m$\upvarepsilon$, linear demodulation results are obtained with a resolution of 0.0688 m$\upvarepsilon$ (2.75\% of the strain range). 
The established linear regression framework can be readily generalized to other types of fiber-optic sensors with similar modulation characteristics (see Supplementary Fig.9--13).

In Fig.~\ref{FBG}(d), we further analyze the strain resolution  across varying measurement ranges, revealing the trade-off between sensing range and resolution, i.e., the system resolution decreases accordingly as the measurement range increases.
This is due to the fact that achieving a balance between high sensitivity to weak signals and linear response to strong signals in photodetectors is inherently challenging. 
Thereby, the ability to discriminate weak signals inevitably decreases when the measurement range is extended. Nevertheless, the measured normalized accuracy remains relatively low and stable across the entire range, indicating robust performance of our system under varying dynamic conditions. 
These experimental results demonstrate the wide applicability of the speckle-based nonlinear encoding and optical computing methods for diverse fiber sensing tasks, which have different requirements in terms of sensing accuracy and measurement range.

\begin{figure}[htbp]
	\centering
	\includegraphics[width=1\linewidth]{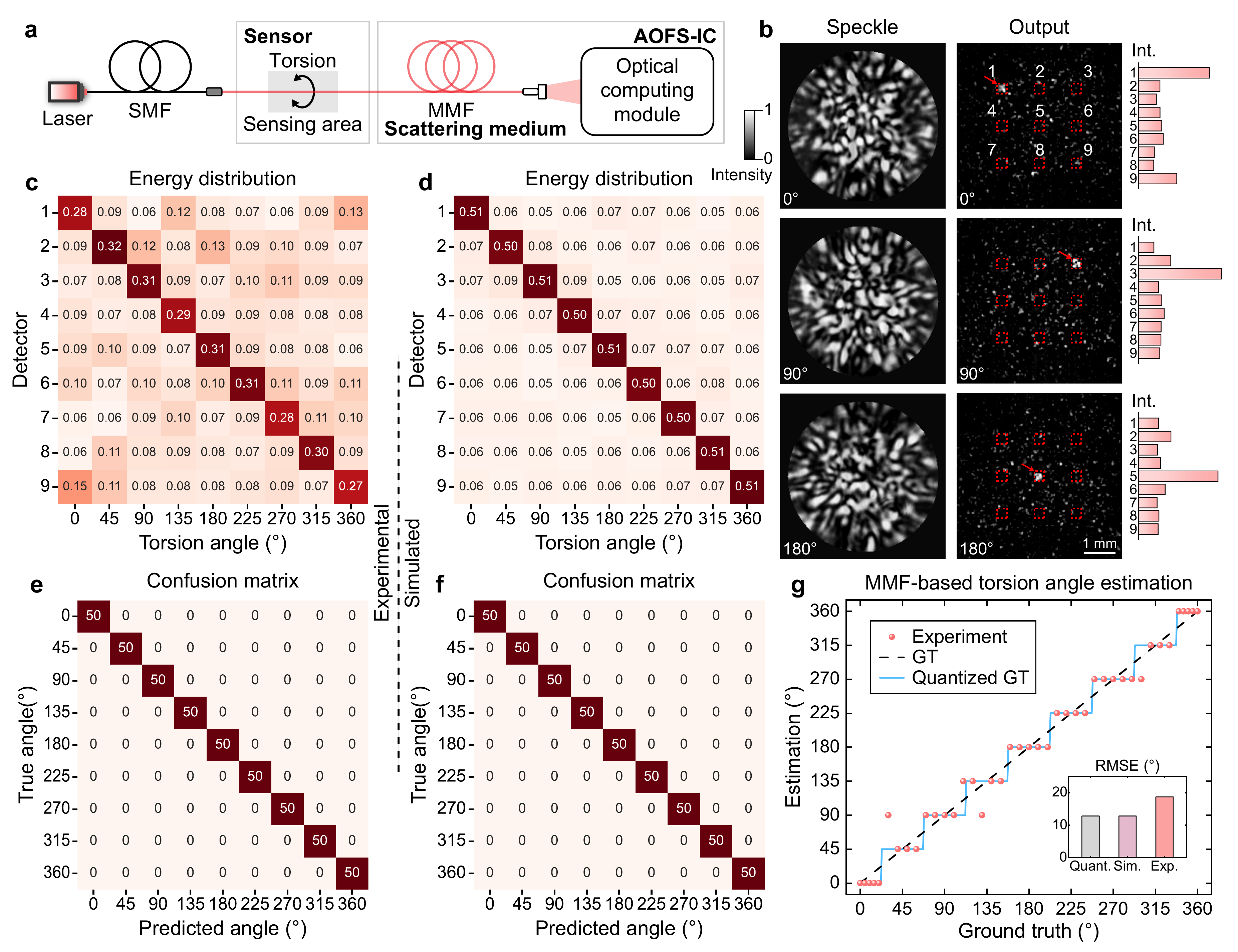}
	\caption{
		\textbf{All-optical classification of discrete torsion states.} \textbf{(a)} Schematic of the experimental setup for torsion sensing. A short segment of MMF serves as the torsion sensor. The set-up of optical computing module is same to Fig.~\ref{FBG}(a). \textbf{(b)} The speckle patterns after passing through the scattering medium (first column), the corresponding output intensity images from the optical diffractive network (second column), and the intuitive intensity distribution (rightmost column) at torsion angles of 0$^\circ$, 90$^\circ$, and 180$^\circ$. The optical computing module exhibits distinct speckle features under varying torsion angles. The detection plane is divided into a 3 $\times$ 3 region array (red dashed boxes) for classification. \textbf{(c--d)} Measured and simulated detector energy distributions across torsion angles from 0$^\circ$ to 360$^\circ$ with 45$^\circ$ spacing. Each torsion state exhibits a dominant intensity localized at a corresponding detector, enabling accurate state identification. \textbf{(e--f)} Confusion matrices based on experimental and simulated predictions, both demonstrating perfect classification accuracy across all tested torsion states. \textbf{(g)}, Quantitative resolution evaluation for all angle classifications from 0$^\circ$ to 360$^\circ$. Quantized GT implies quantized ground truth under perfect classification results. In the subplot, the RMSE of Quantized GT and simulation is 12.81$^\circ$, while the experimental result shows an RMSE of 18.72$^\circ$.
	}
	\label{Class}
\end{figure}

\subsection{All-optical recognition of discrete sensing states}
In certain application scenarios, the object under test may only exhibit a limited number of discrete states, e.g., torsion or bending angles for fixed rotations. 
At this point, the exact magnitude of the measurand becomes less critical, and a simplified state recognition mechanism can be introduced. 
Here, we demonstrate the capability of AOFS-IC to identify discrete torsion states applied to a segment of MMF with a length of 5 cm. As illustrated in Fig.~\ref{Class}(a), MMF is used as torsion sensor, and different torsion angles are realized through controllable rotating device. 
The applied torsion alters the polarization state of the propagating light and introduces additional mode coupling in MMF sensor.
The optical field, whose polarization and intensity distribution are altered, generates distinct speckle patterns after entering the MMF served as the scattering medium in the optical computing module.
The resulting speckle patterns vary significantly with torsion angle, and example speckle images of experimentally captured are shown in Fig.~\ref{Class}(b).
By appropriately adjusting the objective function during ODN training, the speckles corresponding to different torsion angles can be focused onto distinct spatial positions.
The final receiving plane is divided into a 3 $\times$ 3 array of detection regions in the experiment.
Each subregion corresponds to a specific torsion state, where the region receiving the maximum optical energy indicates the estimated torsion angle.
For clarity in presentation, the experimental results in Fig.~\ref{Class}(b) utilize a camera as the detector, which has a limited frame rate.
PD array can be employed in the future to achieve higher sensing performance in real-world scenarios.

To further evaluate performance, experimental and simulated results (see simulation model details in Supplementary Note~1) of detector energy distribution for 9 torsion angles ranging from 0$^\circ$ to 360$^\circ$ are shown in ig.~\ref{Class}(c) and Fig.~\ref{Class}(d), respectively.
For each angle, the highest optical intensity is focused on the right detector, enabling accurate classification. 
The experimental results exhibit a lower signal-to-noise ratio (SNR) compared to the simulation, primarily due to the lack of explicit constraints in the loss function to confine energy within the detector regions (see Methods), as well as a small amount of inherent noise in the physical system. 
Nonetheless, Both the experimental and simulated confusion matrices demonstrate ideal classification accuracy, as shown in Fig.~\ref{Class}(e--f).
All predicted torsion angles perfectly match GT, confirming the capability to reliably distinguish discrete torsion states with 45$^\circ$ resolution steps.

In Fig.~\ref{Class}(g), we quantitatively compare the estimated torsion angles across the entire 0$^\circ$--360$^\circ$ range against GT. 
A quantized ground truth (Quantized GT) is introduced to represent ideal classification at discrete 45$^\circ$ steps. 
The simulated results yield an RMSE of 12.81$^\circ$ (3.56\% of the angular range), identical to the Quantized GT. 
The experimental RMSE is slightly higher at 18.72$^\circ$ (5.20\% of the angular range), primarily due to real-world noise. This performance can be further improved by increasing the number of discrete classification states over the entire angular range, e.g., using a 4 $\times$ 4 detection array to achieve 24$^\circ$ angular steps (see Supplementary Fig.~15).
Our results show that a finite number of classified states can be reliably and stably identified based on the spatial distribution of output optical intensities. 
Moreover, the classification-based optical computing method does not depend on absolute intensity values. 
Its accuracy remains stable even when the optical power is significantly reduced to near the noise floor of PDs (see Supplementary Fig.~16), allowing lower optical power for sensing.

\subsection{Multiplexing of AOFS-IC for multiple positions and multiple measurands}
Benefiting from the scattering medium in AOFS-IC, which is simultaneously sensitive to variations in multiple optical field dimensional parameters, a single speckle pattern can encode diverse sensing information of different physical quantities. 
Further training of the ODN enables the extraction of multiplexed sensing information from the speckle, which makes AOFS-IC has multiplexing capability.

\begin{figure}[htbp]
	\centering
	\includegraphics[width=0.9\linewidth]{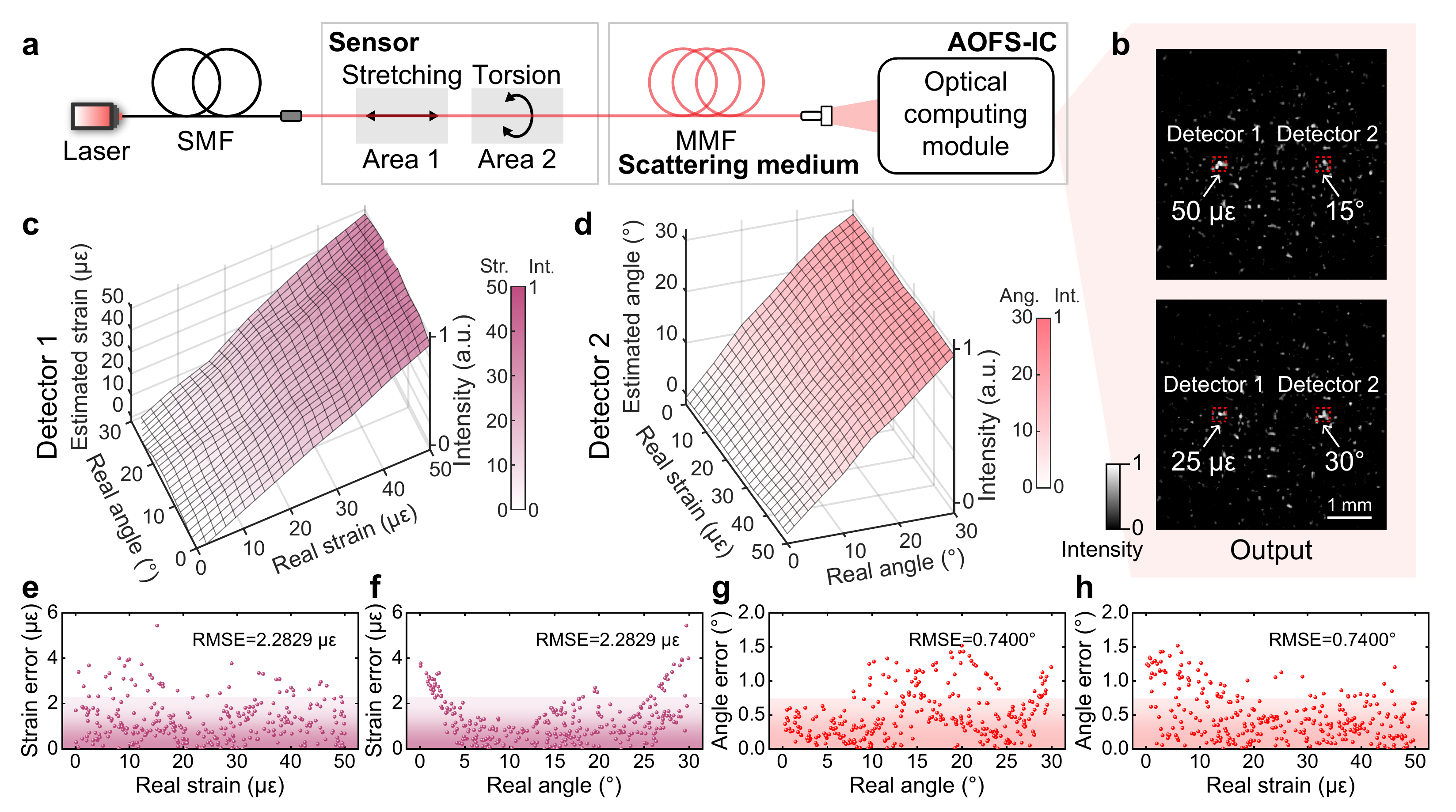}
	\caption{
		\textbf{Simultaneous all-optical sensing of strain and torsion.} \textbf{(a)} Schematic of the experimental setup for dual-parameter sensing, where localized torsion and stretching are applied to different positions of an MMF. The resulting specklegram is processed by an optical computing module to simultaneously extract multiple measurands. \textbf{(b)} Output intensity distributions after passing through the optical computing module corresponding to different combinations of strain and torsion at distinct fiber positions. The area in the detection plane highlighted with a red dashed box indicates the spatial multiplexed readouts. \textbf{(c)} Performance plots for strain (left) and torsion (right) estimation, where the vertical axis indicates the predicted values of the respective measurands under different strain and torsion conditions. Both parameters are accurately and independently recovered, with resolution quantified as 2.2829 $\upmu\upvarepsilon$ for strain and 0.7400$^\circ$ for torsion, validating the capacity of the proposed framework to resolve spatially distributed and cross-sensitive measurands. \textbf{(e--f)} Strain prediction errors under different stretch and torsion conditions, respectively. \textbf{(g--h)} Torsion angle prediction errors under different torsion and stretch conditions, respectively. The purple and red areas indicate errors below the RMSE threshold, reflecting the measurement resolution of the system.
	} 
	\label{Multi}
\end{figure}

An experiment is carried out to verify the multiplexing capability of AOFS-IC, and the experimental setup is shown in Fig.~\ref{Multi}(a).
MMF is selected as the sensor because perturbations at different positions or types induce distinct light field evolutions, thereby generating unique intensity distributions. 
This allows simultaneous application of both torsional and stretch perturbations at different positions along the MMF, which makes a single MMF serves as two distinct sensors: Sensor 1 for torsion and Sensor 2 for stretch.
Notably, while perturbations on both sensors manifest as intensity distribution variations, the speckle patterns generated through high-dimensional nonlinear projections in the scattering medium remain capable of effectively discriminating their respective perturbations (see Supplementary Note~4).
During the training of the ODN, the loss function is extended in dimensionality (see Methods) to suppress mutual interference among different measured quantities, thereby ensuring reliable demodulation performance. 

The trained ODN achieves parallel decoding of speckle patterns from both sensors and focuses their respective sensing information onto spatially separated regions, with the optical intensities in these distinct regions quantitatively representing strain and torsion angle measurements, respectively, as shown in Fig.~\ref{Multi}(b).
The intensities in these regions change synchronously with variations in the input measurands, validating the feasibility of the proposed spatially multiplexed readout mechanism.
Furthermore, Fig.~\ref{Multi}(c) shows the predicted strain values under varying torsion and strain conditions, while Fig.~\ref{Multi}(d) illustrates the predicted torsion angles under the same input conditions.
The strain and torsion prediction errors across different input conditions are summarized in Fig.~\ref{Multi}(e--h). Specifically, Fig.~\ref{Multi}(e) and (f) show the strain estimation errors under all combinations of strain and torsion angle, respectively, while Figs.~\ref{Multi}(g) and (h) display the torsion angle estimation errors under the same parameters.
In all cases, the prediction errors remain at a certain range, indicating reliable decoupling of the two sensing parameters in our system.

The RMSEs of the estimated strain and torsion angle are 2.2829 $\upmu\upvarepsilon$ (4.58\% of the strain range) and 0.7400$^\circ$ (2.47\% of the angular range), respectively. 
For reference, the calibration setup used to generate training data has a strain accuracy of 2 $\upmu\upvarepsilon$ and an angular accuracy of 0.5$^\circ$. 
The experimental results close to these baselines further confirm the high efficiency in multi-parameter decoupling and accurate demodulation. The potential for expanding to more multiplexed sensing locations is additionally confirmed through simulations (see Supplementary Fig.~17).

\subsection{High-speed and high-accuracy all-optical sensing}
Since AOFS-IC acquires sensing information by detecting optical intensities at specific spatial positions, it does not need to use a camera to obtain the intensity distribution of the whole detection plane.
Importantly, inherent low-latency advantage of AOFS-IC would be degraded by camera frame rate limitations.
In the aforementioned experiments, the camera was employed primarily to facilitate both result visualization and multiplexing validation.
AOFS-IC can operate with a single PD as the detector, leveraging inherent advantages of PD to achieve higher sensing performance, such as high speed and high sensitivity.

\begin{figure}[htbp]
	\centering
	\includegraphics[width=1\linewidth]{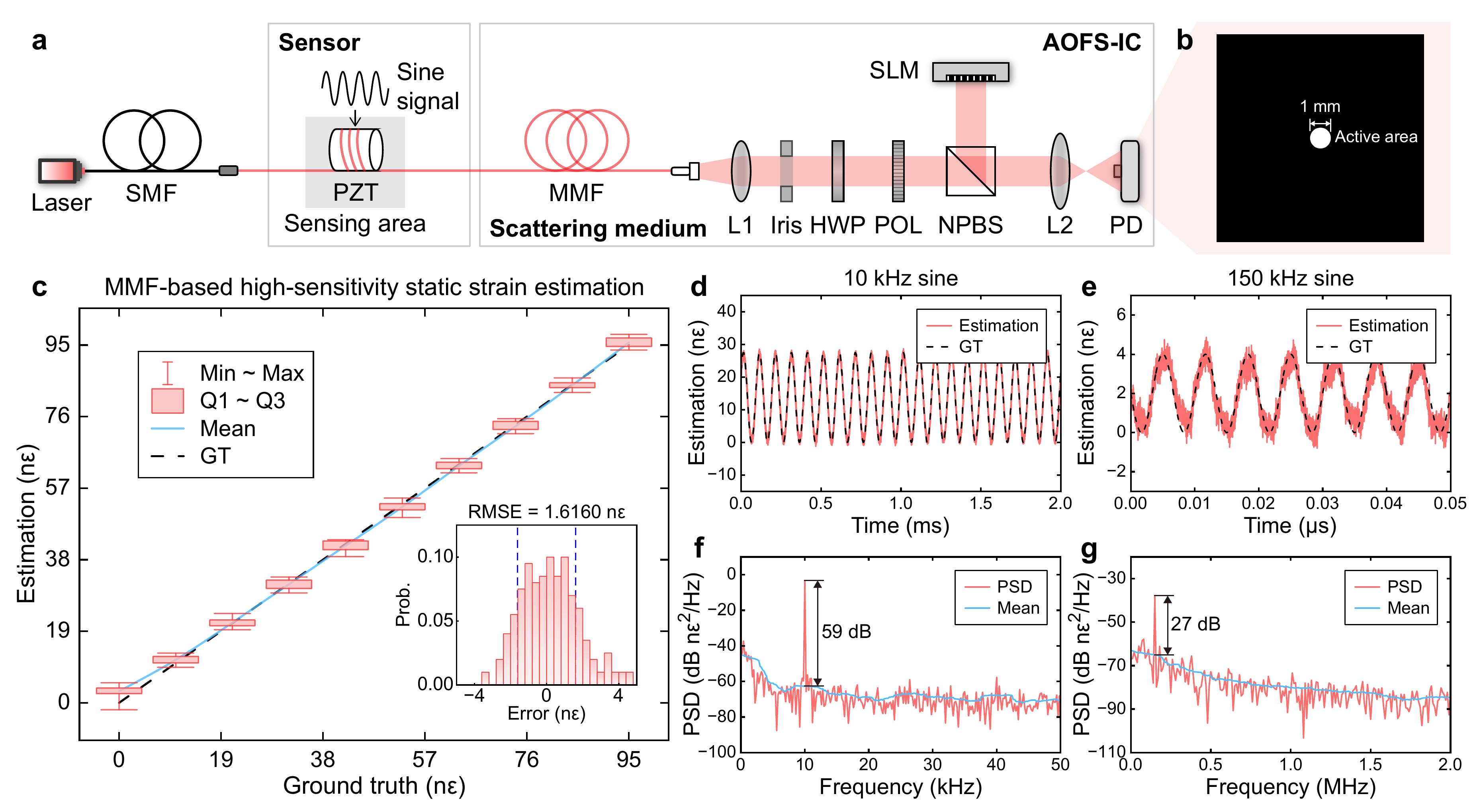}
	\caption{
		\textbf{High-speed and high-accuracy all-optical strain sensing.} \textbf{(a)} Experimental setup for high-speed sensing with a single PD. A segment of MMF sensor is coiled around a PZT to apply dynamic strain. \textbf{(b)} Relative size comparison between the designed speckle intensity region and the effective area of the PD. \textbf{(c)} Performance evaluation of static strain sensing, showing a resolution of 1.6160 n$\upvarepsilon$ over a 95 n$\upvarepsilon$ range, confirming high accuracy signal measurement. \textbf{(d--e)} Time-domain reconstruction results for dynamic strain signals at 10~kHz and 150~kHz. The 10~kHz waveform shows excellent agreement with GT, while the 150~kHz signal exhibits waveform distortion due to PD bandwidth limitations, though its frequency is still correctly retrieved. \textbf{(f--g)} Power spectral density (PSD) analysis of the reconstructed signals at 10~kHz and 150~kHz, showing SNRs of 59~dB and 27~dB, respectively. The reduced SNR at higher frequency further highlight the PD bandwidth constraint.
	} 
	\label{High}
\end{figure}

The high-accuracy strain measurement experimental device using a single PD is shown in Fig.~\ref {High} (a).
A 5-meter MMF segment sensor is coiled around a piezoelectric ceramic transducer (PZT), where strain is induced by electrically driving the PZT. 
From Fig.~\ref{High}(b), the PD features a larger effective receiving area compared to the single camera pixel.
As shown in Fig.~\ref{High}(c), the proposed system achieves an RMSE of 1.6160 n$\upvarepsilon$ within a 95 n$\upvarepsilon$ measurement range (1.70\% of the strain range), demonstrating high sensitivity and accuracy in low-amplitude signal detection. 
This exceptional resolution stems from the inherent responsiveness of MMFs to minute length variations, enabling precise detection of weak strain signals.
The inset histogram of Fig.~\ref{High}(c) illustrates a narrow distribution of estimation errors, indicating excellent measurement repeatability and noise robustness.

To evaluate the sensing performance of high-frequency varying signals, sinusoidal voltages are applied to the PZT to generate dynamic strain signals. 
Fig.~\ref{High}(d--g) shows time-domain and frequency-domain measurement results of 10 kHz and 150 kHz signals. 
The 10 kHz signal is accurately recovered with an SNR of 59 dB, while the 150 kHz signal remains identifiable with a reduced SNR of 27 dB, limited by the PD bandwidth (90 kHz). 
Furthermore, the noise floor of PSD in our system reaches 69 f$\upvarepsilon/\mathrm{\sqrt{Hz}}$ over the 0--5 MHz bandwidth, indicating its potential for ultra-sensitive detection. 
These results validate the proposed PD-based optical computational sensing scheme as a high-speed, high-fidelity sensing solution, making it suitable for applications such as real-time weak signal monitoring.

\subsection{AOFS-IC for robotic arm monitoring and control}

Modern robotic arms and intelligent robotic systems often require multi-parameter and multi-degree-of-freedom (DOF) sensing to ensure precise motion control, operational safety, and adaptability in complex environments. 
These applications pose critical demands on the sensing system, including real-time responsiveness, resistance to electromagnetic interference, and ease of deployment. 
In addition, achieving high-performance sensing with minimal electronic computation and wiring complexity is essential for scalable sensor integration, particularly in systems with numerous joints or in wearable robotics. 
AOFS-IC offers an attractive solution due to its low-latency, low-power, and electronics-free signal processing capabilities. 
By performing optical-domain information computation in the manner of edge computing, AOFS-IC can perform real-time optical-domain information processing without consuming the central computing resources of the robot. 
This makes it highly suitable for distributed, embedded sensing on complex robotic platforms.

\begin{figure}[htbp]
	\centering
	\includegraphics[width=1\linewidth]{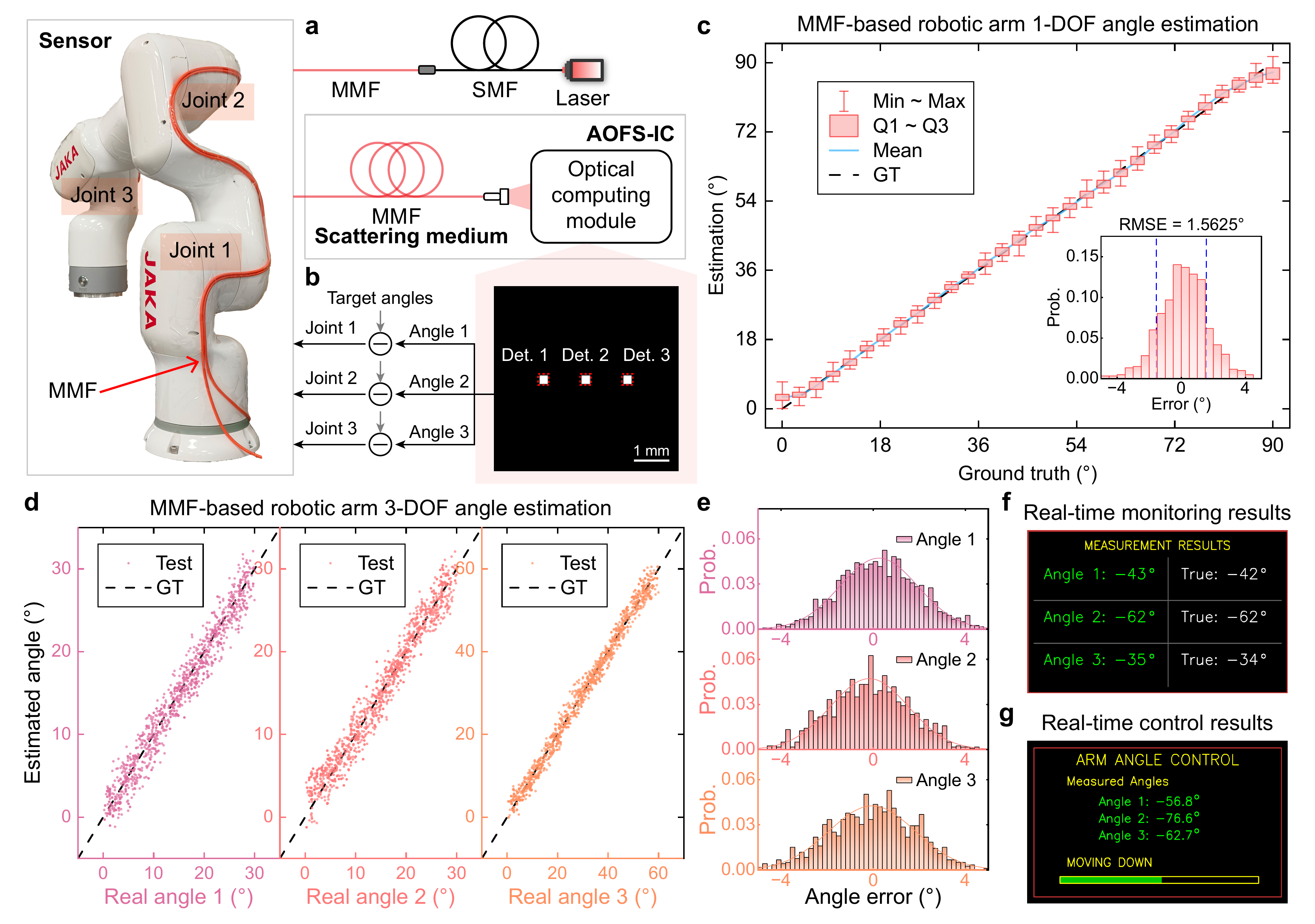}
	\caption{
		\textbf{Multi-degree-of-freedom (DOF) joint angle estimation on a robotic arm using AOFS-IC.} \textbf{(a)} Schematic of the experimental setup and fiber layout for real-time joint angle monitoring. A single MMF is externally routed to cover each rotating joint (Joint 1--3), enabling real-time monitoring of corresponding angles (Angle 1--3) via bending-induced mode coupling.
		\textbf{(b)} AOFS-IC detection plane and generated feedback control signals. Sensing signals are processed and spatially demultiplexed by AOFS-IC, with outputs captured by Det. 1--3.
		\textbf{(c)} Single-DOF angle estimation results. The predicted angles closely match GT with RMSE of 1.5625$^\circ$. Inset shows the corresponding error distribution.
		\textbf{(d)} Multi-DOF estimation performance for Angles 1--3, with calibration ranges of 0--30$^\circ$ for Angles 1 and 2, and 0--60$^\circ$ for Angle 3. All joints show strong linearity and accuracy.
		\textbf{(e)} Angle error histograms for the three joints, with RMSEs of 1.7071$^\circ$, 1.7003$^\circ$, and 1.8755$^\circ$, respectively, confirming high accuracy and repeatability.
		\textbf{(f–g)} Representative user interfaces for real-time joint angle readout (f) and AOFS-IC-driven closed-loop control (g).
	} 
	\label{Arm}
\end{figure}

Fig.~\ref{Arm} illustrates a proof-of-concept experiment demonstrating real-time joint angle monitoring on a 3-DOF industrial robotic arm using AOFS-IC. 
As shown in Fig.~\ref{Arm}(a), a single MMF is tightly routed around the outer surfaces of the robotic arm to cover three rotating joints (Joint 1--3), corresponding to Angles 1--3. 
The MMF shape bending generated by joint rotation induces mode coupling, thereby modulating the transmitted optical signals. 
These modulated signals are then demultiplexed and decoded by the AOFS-IC to extract the encoded motion information. 
Fig.~\ref{Arm}(b) shows the detector distribution of the demodulation results of AOFS-IC and the feedback control process of the robotic arm joint. 
The inferred joint angles are mapped to spatially separated regions on the output plane, where they are read by three detectors (Det. 1--3). 
A camera is currently used for validation despite its frame rate limitations, but future implementations could employ a high-speed PD array to further reduce latency.

Fig.~\ref{Arm}(c) presents the estimation results for a single joint (1-DOF), showing a strong agreement between predicted and true angles across a 90$^\circ$ range, with an RMSE of 1.5625$^\circ$. 
The residual error primarily arises from subtle inconsistencies in fiber deformation---such as friction and surface tension effects---introduced by external mounting, which cause slight deviations in the optical response under identical rotation angles. 
We anticipate that embedding the fiber within the robotic arm structure could mitigate these effects and further improve accuracy. 
To validate multi-joint sensing capability, simultaneous 3-DOF measurements are conducted. Fig.~\ref{Arm}(d) shows scatter plots comparing estimated and actual angles for each joint. The calibration range is set to 0--30$^\circ$ for Joints 1 and 2, and 0--60$^\circ$ for Joint 3. 
All joints exhibit excellent linear correlation between prediction and GT. 
Corresponding error histograms in Fig.~\ref{Arm}(e) confirm that estimation errors are centered around zero and approximately follow Gaussian distributions, demonstrating robust and repeatable decoding. 
The RMSEs for synchronous demodulation of Angles 1, 2, and 3 are 1.7071$^\circ$, 1.7003$^\circ$, and 1.8755$^\circ$, respectively.

In addition to performing monitoring task, AOFS-IC also exhibits the potential for active control of robotic systems, owing to its reconfigurable optical network. 
By training the internal optical pathways to learn mappings between sensed signals and control commands, AOFS-IC can be extended to perform robot closed-loop control tasks. 
Benefiting from the inherently fast response of optical signal propagation, the system can generate control outputs with minimal delay. 
This allows AOFS-IC not only to perform real-time state monitoring (Fig.~\ref{Arm}f), but also to support basic feedback control (Fig.~\ref{Arm}g), both of which are extracted from our live demonstration videos (Supplementary Video 1--2).

\section{Discussion}

The performance of AOFS-IC primarily relies on the configuration of the scattering medium and ODN. 
The scattering medium functions by employing nonlinear high-dimensional projection to transform subtle variations in optical field dimensional parameters induced by the measured physical quantity into distinguishable speckles. 
Due to greater numbers of interference paths and larger optical path differences, complex scattering media with larger volumes may generate speckles that are more sensitive to weak perturbations, thereby enabling higher sensing resolution.
In our demonstration experiments, MMF served as the scattering medium, and it is validated that MMF supporting more guided modes or with longer length achieve superior spectral resolution (see Supplementary Fig.~6--7). 
Generally, sensing resolution is inversely proportional to MMF length, which has also been validated in speckle-based spectrometers\cite{redding2013all, wan2021wavemeter, wan2021review, wan2023reconstructive}.
Although longer MMF offer high sensing resolution, it is more susceptible to external disturbances, compromising system stability.
In contrast, our system maintains high stability over several hours, making it suitable for practical measurements.
Thus, practical application requirements must be carefully weighed to select an appropriate scattering medium. 
When alternative scattering medium is employed in AOFS-IC, the trade-off between sensing performance and stability resembles that of MMF. 
Future efforts may focus on designing miniaturized, complex on-chip scattering medium to achieve high-performance sensing while ensuring long-term stability.

The ODN functions to demodulate the high-dimensional speckle signal generated by the scattering medium in the optical domain, establishing a mapping between the output light intensity and the measured physical quantity. 
Therefore, the capability of ODN to process high-dimensional complex signals is crucial. 
In our experiments, a single-layer optical diffractive network utilizing a SLM has been proved effective. 
Meanwhile, the comparison with simulated diffractive neural networks and Unet demonstrates that our method achieves performance comparable to digital approaches (see Supplementary Fig.~14).
Subsequently, replacing the SLM with a phase diffractive plate could eliminate energy consumption in the optical computing module of AOFS-IC.
Theoretically, increasing the number of optical diffractive network layers and incorporating optical nonlinear functions could further enhance the signal demodulation capability, enabling high-quality sensing measurements for multiplexing of larger number of optical fiber sensors.

The choice of detector type impacts the measurement performance of the final system, a common principle applicable to OFS systems of any architecture. 
The proposed AOFS-IC offers the advantage of compatibility with various detectors. 
When using a single PD, AOFS-IC is better suited for high-speed, high-sensitivity sensing. 
Given the sufficiently low latency of optical computing module in AOFS-IC, the final sensing speed directly matches the response time of the PD.
For detector arrays or cameras, AOFS-IC can leverage spatial multiplexing to achieve the multiplexing of multiple sensors.
Therefore, selecting the appropriate detector is essential to meet the requirements of different application scenarios.

In conclusion, we demonstrate an all-optical fiber sensing architecture with in-sensor computing (AOFS-IC) for realization of light-speed and low-power optical fiber sensing.
Compared to conventional OFS architectures, AOFS-IC achieves signal demodulation entirely in the optical domain with minimal processing latency, which allows for direct quantification of the target physical quantity through detected intensity.
Notably, no optoelectronic conversion occurs prior to obtaining the detected intensity, and the entire measurement process operates without requiring any electronic computational hardwares or resources.
AOFS-IC has been experimentally validated across multiple fiber sensor types, such as FBG, MMF, and SMF, to measure various physical quantities, which alter optical wavelength, polarization state, and intensity distribution in fiber sensor.
Moreover, AOFS-IC achieves simultaneous demodulation of multiple fiber-optic sensors through spatial multiplexing method.
These characteristics suggest AOFS-IC as a promising solution for applications in densely deployed optical fiber sensors.
We demonstrate AOFS-IC based robotic arm posture monitoring that achieves accurate sensing of 3-DOF without utilizing any electronic hardware or computational resources.
With flexible scalability and high performance, our work may inspire next-generation optical-computing-integrated OFS while creating novel opportunities for OFS applications.

\section{Methods}

\subsection{Experimental system}

Our experimental system is built from commercially available optoelectronic devices, forming an optical fiber sensing platform based on all-optical computing. For example, the FBG sensing system employ an ASE light source (AEDFA-23-B-FA) as a broadband energy input to provide the excitation signals for FBG in the 1550 nm band. Other sensing systems that do not require broadband light are fed with a 1550 nm single-frequency continuous wave laser (NKT Photonics). 
A segment of the MMF is used for sensing applications across a wide range of systems, while remaining segment is stored in a temperature controlled box as the scattering medium.
The used MMF feature a 105/125 $\upmu$m core/cladding diameter with step-index profile and 0.22 numerical aperture.
Standard SMF with a 9 $\upmu$m core diameter is also configured in the system to support essential link connections or other sensing tasks.

In the spatial optical computing system, the light output from the MMF is collimated by a high numerical aperture fiber collimator (F950FC-C, Thorlabs), then rotated in the direction of polarization by a zero-level half-wave plate (WPH05ME-1550, Thorlabs), then passes through a zero-order half-wave plate (WPH05ME-1550, Thorlabs) to rotate its polarization direction. A linear polarizer (LPNIR050, Thorlabs) is subsequently used to extract a single polarization state, enabling efficient modulation by the following SLM (PLUTO, Holoeye). The SLM is a reflective phase modulator operating at 1550 nm with 1920 $\times$ 1080 pixels and a linear 2$\uppi$ phase response, capable of applying high-fidelity phase modulation to the incident light field. The reflected light field is reflected into the detection path by a non-polarizing beam splitter (OQPS25.4-1550T5). At the receiving end, a lens with an 80 mm focal length (AC508-080-C, Thorlabs) adjusts the speckle size to meet various spatial multiplexing requirements. The speckle patterns are recorded by an 8-bit infrared camera with a resolution of 256 $\times$ 320 pixels (Bobcat XC251, Xenics). For high-speed reception schemes, light intensity is emitted to an InGaAs PD (PDA10CS(-EC), Thorlabs) with switchable gain, and the resulting electrical signals are recorded using a high-definition oscilloscope (DSOS204A, Keysight), enabling high-speed readout of specific channels.

The measurand is controlled online by programmable devices to provide in situ training datasets. The linear translation stage (DDS220/M, Thorlabs) provides micrometer-scale movement to apply strain and deformation, and the rotation mount (ELL14K, Thorlabs) can apply a determinable angle of torsion. In addition, the RF signal generator (AFG3252, Tektronix) drives the PZT with a voltage to generate the high-frequency vibration signals. The industrial robotic arm (MiniCobo, JAKA) used in the experiment can be controlled online and provide position feedback. During training, the computer triggers frame updates of the SLM and PDs upon changes in the target measurands. In the sensing stage, the system operates autonomously by loading the trained SLM weights, without requiring computer intervention.

\subsection{Training method of ODN}
In the calibration process of AOFS-IC, a mapping from the measurand to the output optical intensity at the receiving end needs to be established. This mapping implicitly integrates the sophisticated physical processes of optical signals, such as optical fiber response to the measurand, nonlinear coding of MMF, spatial light diffraction effect, and phase modulation by SLM. Among them, the phase modulation matrix of SLM serves as the trainable parameter in the calibration process. We adopt an end-to-end training strategy that eliminates the need to explicitly model or individually optimize these intermediate physical processes, simplifying the system calibration process.

Although many state-of-the-art gradient descent algorithms have been developed to support error backpropagation during training\cite{zhou2021large, spall2022hybrid, zheng2023dual}, most of them rely heavily on the accurate simulation of the actual optical pathway by the optical physics model. We employ a nonlinear optimization method based on the genetic algorithm\cite{michalewicz1996genetic} for training on optical systems (see Supplementary Note~7), suitable for rapid deployment and proof of concept. The algorithm we designed allows in-situ training without relying on a physical model, eliminating the need for precise optical alignment of the system.

To estimate the target measurand $M_\mathrm{est}$, we linearly regress the normalized intensity $I_{\mathrm{norm}}$ within the designed region as:
\begin{equation}
M_{\mathrm{est}}= (M_{\mathrm{max}} - M_{\mathrm{min}}) I_{\mathrm{norm} } + M_{\mathrm{min}},
\end{equation}
where $\left [ M_{\mathrm{min}}, M_{\mathrm{max}} \right ]$ denotes the calibration range of the measurand. In the case of single measurand regression, the loss function is constructed based on the RMSE between the predicted intensity and the ground truth $I^\mathrm{obj}$, defined as:
\begin{equation}
Loss_{1}(I, I^{\mathrm{obj}}) = \sqrt{ \frac{1}{N} \sum_{k} (I_{k} - I_{k}^{\mathrm{obj} })^2 },
\end{equation}
Here, $k$ indexes the $k$-th training sample, and $N$ is the total number of training samples. For multi-dimensional regression for multiple positions and multiple measurands, the loss function is extended as:
\begin{equation}
Loss_{2}(\boldsymbol{I}, \boldsymbol{I}^{\mathrm{obj}}) = \sqrt{ \frac{1}{NM} \sum_{k} \left \| \boldsymbol{I} _{k} - \boldsymbol{I} _{k}^{\mathrm{obj} } \right \|_2^2 },
\end{equation}
where $\boldsymbol{I} = (\boldsymbol{I}_{1},\boldsymbol{I}_{2},\dots ,\boldsymbol{I}_{M})^\mathsf{T}$ denotes the optical intensity vector corresponding to $M$ sensing locations or measurands, and $\left \| \cdot \right \| _2$ indicates the Euclidean norm. In classification-based sensing, the region of maximum light intensity identifies the predicted class. If $I_{k,l}$ represents the light intensity in the $l$-th region at the $k$-th class of actual measurand, the classification loss can be defined as:
\begin{equation}
Loss_{3}(I, I^{\mathrm{obj}}) = \sqrt{ \frac{1}{N^2} \sum_{k} \sum_{l} (I_{k,l} - I_{k,l}^{\mathrm{obj} })^2 } ,\ 
I_{k,l}^{\mathrm{obj} }=\begin{cases}
	1,\ k=l \\
	0,\ k\neq l
\end{cases}.
\end{equation}

\subsection{Generalized sensing capability}
To validate the generalization capability of the proposed AOFS-IC in fiber-optic sensing, we demonstrate its applicability across various sensing configurations (see Supplementary Note~9). 
Based on the proposed basic scheme, AOFS-IC can detect the polarization rotation caused by torsion in SMFs and thus perform angle sensing based on the polarization change. 
Furthermore, MMFs are intrinsically sensitive to a wide range of shape perturbations, including radial deformation, axial stretching, angular torsion and bending. 
Several of these sensing modalities have been illustrated, while others are detailed in Supplementary Note~3. Notably, our speckle-based optical computing demodulation system can also be extended to function as a spectrometer or polarization analyzer (see Supplementary Fig.~17--18).

\subsection{Sensing and computing time analysis}
AOFS-IC implements a fully optical sensing and computation paradigm, eliminating the need for electronic computer-based storage and digital processing. Since optical signals propagate at the speed of light within the fiber, the transmission latency is primarily determined by the propagation time from the sensing region to the demodulation module, given by $\tau_{\mathrm{trans}}=nL/c$, where $n$ is the refractive index, $L$ is the fiber length, and $c$ is the speed of light. An additional segment of MMF is inserted before the optical computing demodulator for nonlinear encoding, contributing to propagation delay. In our general experimental validation, the MMF used for high-dimensional mapping can be reduced to 0.5 meters or even shorter (see details in Supplementary Note~8), corresponding to an encoding delay time of $\tau_{\mathrm{encode}} = 2.44$ ns. The all-optical computing process, based on spatial diffraction and phase modulation, introduces negligible demodulation latency ($\tau_{\mathrm{compu}} = 0.08$ ns) due to its short free-space optical path. In the demodulation process of AOFS-IC, the total response time from instantaneous measurement changes to demodulation results is estimated to be \textless 3 ns. This represents a significant improvement over traditional fiber optic sensors, which typically exhibit demodulation speeds of at least microseconds.

\subsection{Noise analysis}
In our experiment, system noise arises from multiple sources. 
A primary contributor is the temporal fluctuation of MMF-generated speckle patterns, quantified by the correlation coefficient between successive speckle frames. 
Over a 24-hour continuous camera observation, the speckle correlation gradually decreased to 0.98 (see Supplementary Fig.~5), indicating that the speckle patterns remain highly consistent over time, ensuring robust sensing performance. 
This fluctuation mainly results from slow environmental thermal drift, which can be further mitigated through enhanced temperature control. 
Other noise contributions include laser frequency drift and phase noise, which are amplified through nonlinear mode interfere in the MMF. 
On the detection side, noise sources such as shot noise, dark current noise, and readout noise become particularly relevant, epscially under low-power conditions. 
Overall, the system achieves a favorable balance between power consumption and noise control, offering strong demodulation stability.

\bigskip

\section*{Acknowledgements}
This work is financially supported by National Natural Science Foundation of China (NSFC) under Grant No. 62405178, 62435004.
We would like to express our sincere gratitude to Professor Weichao Guo and Professor Xiangyang Zhu from Shanghai Jiao Tong University for providing the robotic platforms and for their valuable assistance and guidance.

\section*{Data, Materials, and Code Availability} 
The data that support the findings of this study are available from the corresponding author upon reasonable request.

\section*{Author contributions} 
Y. Wan conceived the idea.
Y. Tao performed the experiments, analyzed the results.
Y. Tao and Y. Wan prepared the manuscript. 
Z. Long, J. Du, and W. Zhang provided experimental equipment and assistance.
Z. He and Y. Wan discussed the work and revised the paper.

\section*{Conflict of interest} 

The authors declare no competing interests.

\bibliography{ref}   
\bibliographystyle{ieeetr}

\end{spacing}

\end{document}